\documentclass[letterpaper,9pt]{IEEEtran}
\IEEEoverridecommandlockouts
\usepackage{color}
\usepackage{enumerate,graphicx,epstopdf}
\usepackage{amsmath,amssymb,bm,amsthm}
\usepackage{cite}
\usepackage{mathtools}
\usepackage{array}
\usepackage{algpseudocode,algorithm,algorithmicx}
\usepackage{color}
\usepackage{booktabs}
\usepackage[hyphens]{url}
\usepackage{lipsum}
\usepackage[dvipsnames]{xcolor}
\usepackage[deletedmarkup=sout,authormarkup=superscript]{changes}
\usepackage[normalem]{ulem}
\usepackage{flushend}

\usepackage[hidelinks]{hyperref}
\usepackage[english]{babel} 

\usepackage[makeroom]{cancel} 

\usepackage[overload]{empheq}
\usepackage[belowskip=-10pt,aboveskip=0pt]{caption}
\usepackage{subcaption}
\usepackage{multicol}

\let\labelindent\relax
\usepackage{enumitem} 

\newtheorem{ass}{Assumption}
\newtheorem{rmk}{Remark}

\newtheorem{prp}{Proposition}
\newtheorem{cor}{Corollary}
\newtheorem{thm}{Theorem}

\newcommand{\R}{\mathbb{R}}

\newcommand{\mc}[1]{\mathcal{#1}}
\newcommand{\bb}[1]{\mathbb{#1}}

\newcommand{\col}{\mathrm{col}}
\newcommand{\mbf}[1]{\mathbf{#1}}

\newcommand{\x}{\mathbf{x}}

\graphicspath{{figs/}}

\definechangesauthor[name={Sophie}, color=NavyBlue]{SH}
\definechangesauthor[name={Dominic}, color=RedViolet]{DLM}\definechangesauthor[name={Bepe}, color=OliveGreen]{GB}

\begin{document}
\makeatletter
\newcommand{\printfnsymbol}[1]{%
  \textsuperscript{\@fnsymbol{#1}}%
}
\makeatother
\title{Stability Certificates for Receding Horizon Games}

\author{Sophie Hall, Giuseppe Belgioioso, Florian D\"{o}rfler, Dominic Liao-McPherson
\thanks{Sophie Hall, Giuseppe Belgioioso, and Florian D\"{o}rfler are with the Automatic Control Laboratory, ETH Z\"{u}rich, Switzerland.
(emails: \texttt{\{shall, gbelgioioso, dorfler\}@ethz.ch}. Dominic Liao-McPherson is with the University of British Columbia, Vancouver, Canada, (e-mail:
\texttt{dliaomcp@mech.ubc.ca}). This work is supported by the SNSF via NCCR Automation (Grant Number 180545). 
}
}
\pagestyle{empty}
\maketitle
\thispagestyle{empty}
\begin{abstract}
Game-theoretic MPC (or Receding Horizon Games) is an emerging control methodology for multi-agent systems that generates control actions by solving a dynamic game with coupling constraints in a receding-horizon fashion. 
This control paradigm has recently received increasing attention in various application fields, including robotics, autonomous driving, traffic networks, and energy grids, due to its ability to model the competitive nature of self-interested agents with shared resources while incorporating future predictions, dynamic models, and constraints into the decision-making process.
In this work, we present the first formal stability analysis based on dissipativity and monotone operator theory that is valid also for non-potential games. Specifically, we derive LMI-based certificates that ensure asymptotic stability and are numerically verifiable. Moreover, we show that, if the agents have decoupled dynamics, the numerical verification can be performed in a scalable manner. Finally, we present tuning guidelines for the agents' cost function weights to fulfill the certificates and, thus, ensure stability. 
\end{abstract}


\section{Introduction}
\label{sec:introduction}


Our modern society is underpinned by a multitude of shared infrastructure and resources (e.g., traffic networks, power grids, and fish stocks). These systems and resources need to be carefully managed; if left unchecked, selfish behavior by users can lead to severe societal losses: selfish routing decisions lead to traffic congestion~\cite{roughgarden2002how}, selfish energy consumption causes blackouts~\cite{matthewman2014blackouts}, and overfishing leads to the collapse of fish stocks~\cite{gordon1954economic}. The need to make these critical systems more resilient is driving the development of resource allocation protocols that manage them in a fair and sustainable way.

In many cases these decisions need to be made in real-time through control algorithms. Controlling shared infrastructure and allocating resources in real-time is a challenging problem for multiple reasons: (i)~it requires optimization of general economic cost functions, (ii)~resource availability and prices are time-varying requiring predictions into the future (e.g., fish stock varies with the seasons and electricity prices vary with renewable generation), (iii)~enforcing local and global constraints is necessary as system infrastructure underlies strict operating conditions and shared resources are limited, (iv)~large-scale systems require efficient distributed solution algorithms. One of the only systematic and tractable control frameworks for systems with these properties is Economic Model Predictive Control (EMPC) as it allows for general economic objectives and constraints as well as for distributed implementations~\cite{mueller2017economic}. However, EMPC implicitly assumes that agents accessing the shared resources are willing to cooperate fully to achieve a socially-optimal outcome. Making such an assumption seems unreasonable, it is well established that self-interested agents accessing a shared resource optimize for their own benefit, e.g., see the infamous \textit{tragedy of the commons}~\cite{hardin1968tragedy}, leading to outcomes that are far from socially optimal~\cite{koutsoupias1999worst}.


    



Game theory comes to the rescue as a powerful tool for modeling conflict and cooperation between self-interested decision makers. In particular, network games have been used extensively to study resource allocation problems, see~\cite{belgioioso2022distributed} and references therein. %
However, these works assume the system is in steady-state whereas in reality most engineering and natural systems are governed by dynamics. Neglecting these means that (i) transients are not optimized~\cite{koehler2023transient}, (ii) constraints may be violated, and (iii) solutions may not coincide with steady states of the system~\cite{mazumdar2020gradient}. Moreover, introducing feedback into the system is essential for robustness against disturbances and model mismatch. One approach to do so is to use explicit game-theoretic feedback policies~\cite{nortmann2024nash}, however for the setting with coupling constraints these policies are expensive to compute and only approximate schemes exist to date~\cite{laine2023computation}. Another approach is to solve the game in a receding-horizon fashion, measuring the state and recomputing the open-loop trajectory at every sampling time, \`a la MPC. 



\begin{figure}[t]
\centering
\includegraphics[width=.9\columnwidth]{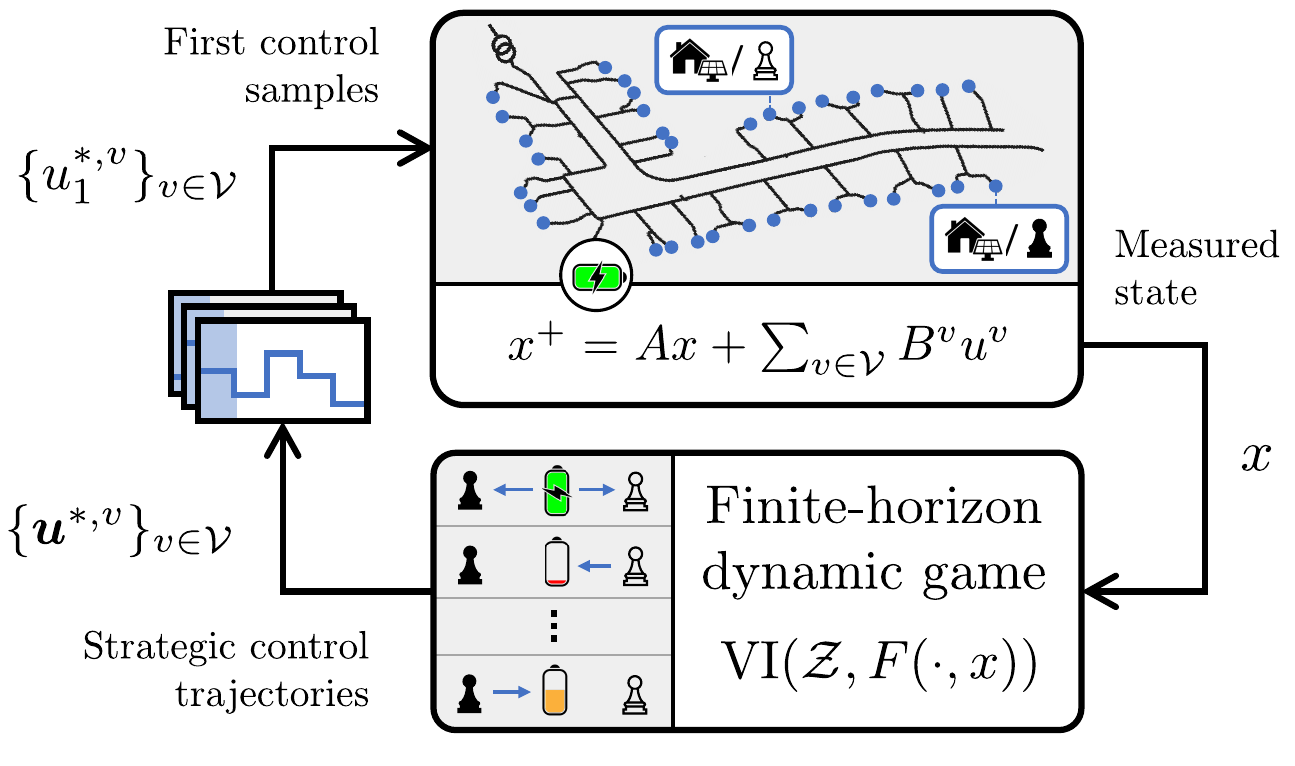}\vspace{0.5em}
\caption{In RHG, control actions are generated by solving a finite-horizon dynamic game, in a receding-horizon fashion.} \label{fig:RHG_scheme}
\end{figure}

The combination of MPC and game theory, also called Receding Horizon Games (RHG)~\cite{hall2022receding} or Game-Theoretic Planning~\cite{spica2020real, wang2021game}, allows handling systems with dynamics, constraints, and self-interested agents, see Figure~\ref{fig:RHG_scheme}.
This control paradigm has already been successfully employed in various engineering applications, including supply chains~\cite{hall2024game},  robotics~\cite{gu2008differential}, autonomous driving~\cite{liniger2020noncooperative, wang2021game, lecleach2022algames}, electric vehicle charging~\cite{mignoni2023distributed}, and smart grids~\cite{paola2018distributed, hall2022receding}. Despite its widespread application, RHG still lacks the rigorous underpinning of MPC, such as stability and recursive feasibility. Yet, for large-scale infrastructures such theoretical guarantees are essential to ensure efficient, safe, and reliable operation at all times.

Up to now such guarantees have only been established for strongly monotone potential games played in receding-horizon~\cite{hall2022receding,benenati2024probabilistic}. Note that for dynamic resource allocation problems potential games are idealized and practically very restrictive as they constrain agent's cost function design and thus cannot model scenarios where agents truly act in their own self-interest. The authors in~\cite{benenati2024probabilistic} discuss a receding-horizon implementation of a game-theoretic traffic routing system and show stability under a strong monotonicity assumption for a potential cost which is proven to be a control Lyapunov function of the closed-loop system. In~\cite{venkat2005stability} a game-theoretic distributed MPC with coupled dynamics and local input constraints is designed. A potential-like global cost function is derived through weighted sums of strongly convex stage costs and closed-loop stability is proven.


In this paper, we leverage monotone operator theory and dissipativity theory to develop the first general stability analysis of RHG. Overall, the paper makes the following three main contributions: 
 \begin{enumerate}
 \item We derive a sufficient condition for the closed-loop stability of RHG controllers that can be numerically verified by solving a linear matrix inequality (LMI). Our certificates cover RHG controllers with strongly monotone games and linear stable dynamics. Moreover, we establish conditions under which these certificates can be computed in a distributed manner.
 \item We derive analytical solutions to the LMIs in a simplified setting and identify key factors affecting the closed-loop stability of RHG controllers. Based on these analytical results, we provide local tuning recommendations for RHG guidelines.
 \item We validate the effectiveness of our stability certificates and tuning guidelines via a case study on a battery charging scenario, demonstrating their practical applicability in real-world settings.
 \end{enumerate}





Our stability analysis is related to the technical analysis in~\cite{fabiani2023incentives}, where linear state feedback matrices $K$ are designed to guide a group of self-interested agents to a co-evolutionary equilibrium. Stability certificates for strongly monotone games are derived by leveraging similar dissipativity tools. However, the focus lies on the design of incentives without predictions, which differs significantly from the perspective of our work. In~\cite{copp2019distributed}, stability is proven for a multi-agent consensus problem under receding-horizon control, where agents solve a min–max optimization to minimize local tracking errors. Due to differing solution concepts, stability results are not comparable.
\section{Preliminaries}

%
\subsubsection{Basic notation} We denote by $\bb{Z}_K= \{0,\dots, K\!-\!1\}$ the sequence of the first $K$ non-negative integers. The set of positive definite symmetric matrices of dimension $n\times n$ is denoted as $\mathbb{S}_{\succ 0}^n$. Given $M$ vectors $u^1, ..., u^M$, we denote  by  $ u = \text{col}(u^v)_{v=1}^M:= [(u^1)^\top, \ldots, (u^M)^\top]^\top$ the stacked vector of vectors $u^v$.
%
%
%
Given $K$ matrices, $\text{blkdiag}(H_i)_{i=1}^K$ denotes the block diagonal matrix with $H_1 , . . . , H_K$ on the main diagonal. The spectral radius of a matrix $H$ is defined as $\rho(H) = \max\{ |\lambda_1|,\dots ,|\lambda_n|\}$ where $\lambda_1,\dots,\lambda_n$ are the eigenvalues of $H$.
%
\subsubsection{Operator Theory} A set-valued operator $T: \R^q \rightrightarrows \R^q$ is ($\mu$-strongly) monotone if $\langle x-y,u - v\rangle  \geq 0,$ ($\geq\mu~\|x-y\|^2$), $\forall x,y \in \R^q$, with $u \in T(x), v \in T(y)$. A single-valued operator $S: \R^q \to \R^q$ is $\beta$-cocoercive if $\langle x-y, S(x) - S(y)\rangle \geq \beta~\| S(x)-S(y)\|^2$ for all $x,y \in \R^q$.

\section{Receding Horizon Games}
\label{sec:RHG}



Consider a linear time-invariant (LTI) system of the form
\begin{align}\label{eq:LTI}
x_{t+1} = A \, x_t + \sum_{v\in \mc{V}} B^v \, u^v_t ,
\end{align}  
 which is controlled by a group of agents $v \in \mc{V} = \{1,\dots, M\}$ where $x_t\in \R^{n_x}$ is the global state at time $t$ and $u_t^v\in \R^{n^v_u}$ is the input of agent $v$. At time step $t$ (or stage), agent $v$ incurs the cost
\begin{align}\label{eq:StageCost}
&\ell^v(x_t,u^v_t, u^{-v}_t)=x_t^\top W^v x_t + (w^v)^\top x_t + \ell_u^v(u_t^v,u_t^{-v}),
\end{align}
where $\ell_u^v: \R^{n_u} \to \R$ is the actuation cost for agent $v$ and $W^v\in \mathbb{S}_{\succ 0}^{n_x} $.
%
This so-called \textit{stage cost} depends on the global state $x_t$, the local control input $u^v_t$, and the input of all other agents. Additionally, agent $v$ must satisfy at all times the following constraints:
\begin{subequations}
    \begin{align}\label{eq:LocalConstr}
u^v_t &\in \mc{U}^v\\
\label{eq:CouplingConstr}
(u_t^v, u_t^{-v}) &\in \mc{C},
\end{align}
\end{subequations}
where~\eqref{eq:LocalConstr} models local input constraints and~\eqref{eq:CouplingConstr} is a joint coupling constraint which models the agent's access to a shared resource. 


Overall, each agent $v \in \mathcal{V}$ aims to minimize its infinite horizon cost while satisfying both local and coupling constraints, namely,
\begin{align}\label{eq:InfiniteProb} v\in \mc{V}: \left\{
\setlength{\arraycolsep}{2pt}%
\begin{array}{r llll}
&&\displaystyle \min_{u^v,\, x}  &\; \displaystyle \sum_{k= 0}^{\infty} \ell^v(x_k,u^v_k,u_k^{-v})\\   
&&\textrm{s.t.} \quad &  x_{k+1} =  A\, x_k + {\textstyle\sum_{v\in \mc{V}}} B^v\, u^v_k, \quad k \in \bb{Z}_{\infty}\\
&&& u_k^v \in \mc{U}^v, \hspace*{9.5em}
k \in \bb{Z}_{\infty}\\
&&&(u_k^v, u^{-v}_k) \in  \mc{C}, 
\hspace*{7.2em}
k \in \bb{Z}_{\infty}
\\ 
&&& x_0 = \mbf{x},\end{array}
\right.
\end{align}
%
where $\mbf{x}$ is the initial state of system~\eqref{eq:LTI}.

However, as~\eqref{eq:InfiniteProb} is generally computationally intractable, we consider instead the finite-horizon version: 
\begin{subequations}
\label{eq:MPCagent}
\begin{empheq}[left=v\in \mc{V}:  \empheqlbrace]{align}
\label{eq:RunningCost}
&&\displaystyle \min_{u^v,\, x}  &\; \displaystyle \sum_{k= 0}^{K-1} \ell^v(x_k, u^v_k,u_k^{-v})\\ \label{eq:Constr1}
&&\textrm{s.t.} \quad &  x_{k+1} =  A\, x_k + {\textstyle\sum_{v\in \mc{V}}} B^v\, u^v_k , \quad  k \in \bb{Z}_{K}\\
\label{eq:Constr2}
&&& u_k^v \in \mc{U}^v,\hspace*{9.5em}
k \in \bb{Z}_{K} \\\label{eq:Constr3}
 &&&(u_k^v, u^{-v}_k) \in  \mc{C}, 
\hspace*{7.2em}
k \in \bb{Z}_{K}
\\ 
&&& x_0 = \mbf{x},
\label{eq:Constr4}
\end{empheq}
\end{subequations}
where $K>1$ is the prediction horizon. Note that we do not include terminal state constraints in the RHG problem~\eqref{eq:MPCagent}, which is commonly done in MPC. These $M$ interdependent optimal control problems (OCPs), parametric in $\mbf{x}$, constitute a non-cooperative generalized game, in which both cost and admissible decision set of each agent depend on others~\cite{facchinei2009generalized}.
To synthesize an RHG feedback law based on the coupled OCPs in~\eqref{eq:MPCagent}, we first define the global decision set
\begin{align}\label{eq:GlobalDecision}
 \mc{Z} = \prod_{v\in\mc{V}} \mc{U}^v~\cap~\mc{C}.
\end{align}
Then, we substitute the dynamics~\eqref{eq:Constr1} and initial condition~\eqref{eq:Constr4} into the stage cost~\eqref{eq:RunningCost} of agent $v$ and denote the resulting cost function as $J^v(u^v, u^{-v}, \mbf{x})$. This leads to the following compact formulation of the game~\eqref{eq:MPCagent}:
\begin{equation} \label{eq:GamePerAgent}
\forall v\in \mc{V}:\quad \left\{
\begin{array}{rl}
 \displaystyle \min_{u^v}& {J^v(u^v, u^{-v}, \mbf{x})}\\
\textrm{s.t.} & (u^v, u^{-v}) \in \mc{Z}.
\end{array}
\right. 
\end{equation}
A set of strategies $\col(u^{*,v})_{v\in\mc{V}}$ that simultaneously solve~\eqref{eq:GamePerAgent} is a generalized Nash equilibrium (GNE), see~\cite{facchinei2009nash} for a formal definition. Among all possible GNEs, we consider the subclass of \textit{variational GNEs} (v-GNEs), which correspond to the solutions of the following generalized equation:
\begin{align}\label{eq:VI}
\text{F}( u^*, \mbf{x}) + \mc{N}_{\mc{Z}}(u^*) \ni 0,
\end{align}
where F$(u, \x) = \text{col}(\nabla_{u^v} J^v(u^v, u^{-v}, \mbf{x}))_{v\in \mc{V}}$ is the pseudo-gradient of~\eqref{eq:GamePerAgent}, and $\mc{N}_{\mc{Z}}$ is the normal cone~\cite[Def.~6.38]{bauschke2017convex} of the global decision set $ \mc{Z}$ in~\eqref{eq:GlobalDecision}. We note that $\mc{N}_{\mc{Z}}$ can be given explicitly for all standard constraint sets. Variational GNEs are often desirable solutions for dynamic resource allocation problems as they satisfy coupling constraints~\eqref{eq:CouplingConstr}, are strategically-(Nash-)stable, i.e., no agent has an incentive to deviate from their agreed upon decision, and are ``economically fair'', since each agent incurs the same marginal loss due to the presence of the coupling constraints~\cite{facchinei2009nash}. 

We define the mapping from initial state $\mbf{x}$ to the solutions of the generalized equation (\ref{eq:VI}), or equivalently, to the v-GNEs of~\eqref{eq:MPCagent}, as
\begin{align} \label{eq:SolMap}
\mc{S}(\mbf{x}) = \{
u ~|~ F(u, \mbf{x})+ \mc{N}_{\mc{Z}}(u) \ni 0 \}.
\end{align} 
In the remainder of this section, we proceed under the assumption that $\mc{S}$ is always well-defined and single-valued. Precise conditions ensuring this will be formalized later in Assumption~\ref{ass:Global}.

To approximate a solution of the infinite-horizon game~\eqref{eq:InfiniteProb}, we recursively solve the finite-horizon version~\eqref{eq:MPCagent}, in a receding-horizon fashion. Specifically, at each sampling time $t$, the agents compute the v-GNE of the game (\ref{eq:MPCagent}) with measured state $\x_t$, i.e., \mbox{$u^* = \text{col}(u_k^*)_{k\in\bb{Z}_{K}}$}, and then apply the first element $u^*_0$ of the optimal control trajectory. This creates an implicit feedback policy $\kappa$ (referred to as the RHG feedback law) defined as
\begin{align}\label{eq:FeedbackLaw}
u^*_t =  \kappa(\x_t)   =  \Xi\, \mc{S}(\x_t),
\end{align}
where $\Xi$ is a selection matrix that extracts the first element of the control sequence of each agent $v\in \mc{V}$, namely, $ u_0^{*,v}$. 
The resulting closed-loop system under the RHG feedback law~\eqref{eq:FeedbackLaw} is
\begin{align}\label{eq:ClosedLoop}
x_{t+1} = A\, x_t + B\,\kappa(\x_t),
\end{align}
where $B = [B^1, \dots,B^M]$ collects the local $B^v$ matrices in~\eqref{eq:LTI}.


\section{Stability Conditions}
In this section, we derive conditions under which the closed-loop system (\ref{eq:ClosedLoop}) is stable. We make the following standing assumptions. 

\begin{ass}\label{ass:Global} The following conditions hold:
\begin{itemize}
\item[(i)] The open-loop dynamics~\eqref{eq:LTI} are stable, i.e., \mbox{$\rho(A)< 1$}.
\item[(ii)] The coupling constraint set $\mc{C}$ and the local sets $\mc{U}^v, \forall v\in \mc{V}$, are closed and convex; the set $\mc{Z}$ is compact and non-empty; the function $\ell_u^v$ is continuous and $\ell_u^v(\cdot,u^{-v})$ is strongly convex and continuously differentiable, for any fixed $u^{-v}$. 
\item[(iii)] The pseudo-gradient $F(\cdot,\mbf{x})$ in~\eqref{eq:VI} is $\mu$-strongly monotone for any fixed $\mbf{x}$.
\end{itemize}
\end{ass} 
Assumption \ref{ass:Global}(i) is widely adopted in dissipativity-based stability analyses, and can be enforced, for example, by pre-stabilizing \eqref{eq:LTI} with a linear feedback gain $K$ such that $\rho(A- \sum_{v\in\mc{V}} B^v K)<1$. 
Assumption~\ref{ass:Global}(ii) aligns with the standard requirements of regularity and convexity of game-theoretic works ~\cite{facchinei2009nash}. Coupled with Assumption~\ref{ass:Global}(iii), these conditions ensure existence and uniqueness of a v-GNE~\cite[Th. 2.3.3]{facchinei2007finite}, as formalized next. Within the classic MPC framework, the latter assumption corresponds to strong convexity of the stage cost.
\begin{prp} \label{prp:Unique}
Under Assumption~\ref{ass:Global}, the solution mapping $\mc{S}(\mbf{x})$ in~\eqref{eq:SolMap} maps to a singleton, for all arguments $\mbf{x} \in \R^{n_x}$.
\end{prp}
\begin{proof}
The normal cone of a convex set $\mc{N}_{\mc{Z}}$ is maximally monotone, and $F(\cdot,\mbf{x})$ is strongly monotone by Assumption~\ref{ass:Global}~(ii). Thus, $\mc{S}(\mbf{x})$ in~\eqref{eq:SolMap} maps to a singleton by ~\cite[Cor. 23.37]{bauschke2017convex}.
\end{proof}

The core idea of our analysis is to view the closed-loop system~\eqref{eq:ClosedLoop} as a feedback interconnection of two dissipative systems, an LTI system and a static nonlinearity.
We leverage known links between monotone operator theory and dissipativity to derive conditions which ensure asymptotic stability of this interconnection~\cite[Def. 2.14]{gruene2011nonlinear}.

By exploiting the separable structure of the stage costs in~\eqref{eq:StageCost}, we can write the pseudo-gradient mapping in~\eqref{eq:PseudoGrad} as 
\begin{align}\label{eq:PseudoGrad}
F(u, \x) = F_u(u) + F_x \x
\end{align}
where the expressions for $F_u(\cdot)$ and $F_x$ are the following
\begin{align}\label{eq:Fu} \nonumber
&F_u(u) =
\left[\begin{smallmatrix}
\partial_{u_0^1} \ell_u^1(u_0^1,u_0^{-1}) \\
\vdots\\
\partial_{u_{K-1}^M} \ell_u^v(u_{K-1}^M,u_{K-1}^{-M})
\end{smallmatrix}\right] \\ 
&+\left[\begin{smallmatrix}
2(\tilde{B}^{1})^\top \tilde{W}^1\tilde{B}^{1}&\hdots&(\tilde{B}^{1})^\top \tilde{W}^1\tilde{B}^{M}\\
 \vdots& \ddots&\vdots\\
 (\tilde{B}^{M})^\top \tilde{W}^{M}\tilde{B}^{M}&\hdots& 2(\tilde{B}^{M})^\top \tilde{W}^{M}\tilde{B}^{M}
\end{smallmatrix}\right]
u
+\left[\begin{smallmatrix}
(\tilde{B}^{1})^\top \tilde{w}^1 \\
\vdots\\
(\tilde{B}^{M})^\top \tilde{w}^M 
\end{smallmatrix}\right]
\end{align}
\begin{align}\label{eq:Fx}
F_x = \left[\begin{smallmatrix}
2(\tilde{B}^{1})^\top \tilde{W}^1 \tilde{A}\\
\vdots\\
2(\tilde{B}^{M})^\top \tilde{W}^M \tilde{A}
\end{smallmatrix}\right],
\end{align}
with $\tilde{A}$ being the free response matrix of the global dynamics~\eqref{eq:LTI}, and $\tilde{B}^{v}$ the impulse response matrix of agent $v$, namely,
\begin{align}\label{eq:ResponseMat}
\tilde{A}=
 \left[\begin{smallmatrix}
  I \\
  A\\
  \vdots\\
  (A)^K
  \end{smallmatrix}\right], \quad \tilde{B}^v =  \left[\begin{smallmatrix}
0 & \cdots & \cdots & 0\\
B^v & 0 & \cdots & 0\\
AB^v & B^v & \cdots & 0\\
\vdots & \ddots & \ddots & 0 \\
(A)^{K-1}B^v & \cdots & AB^v & B^v
\end{smallmatrix}\right],
\end{align}
and, further, $\tilde{W}^v= \text{blkdiag}(W^v, \dots, W^v), \; \tilde{w}^v = \text{col}(w^v, \dots, w^v)$. 
With this result in place, we can define the mapping 
\begin{align}\label{eq:Phi}
\phi(\cdot) = (F_u + \mc{N}_{\mc{Z}})^{-1}(\cdot)
\end{align}
which facilitates deriving a more explicit expression for $\mc S$. The mapping $\phi(\cdot)$ obeys the following technical  properties.
\begin{prp}\label{prp:PhiProperties} The mapping $\phi(\cdot)$ is $\mu$-cocoercive.
\end{prp}
\begin{proof}
This results from $\phi(\cdot)$ being the inverse mapping of a $\mu$-strongly monotone operator ~\cite[Ex. 22.7]{bauschke2017convex}.
\end{proof}

\begin{figure}
\begin{center} 
\includegraphics[width=0.98\columnwidth]{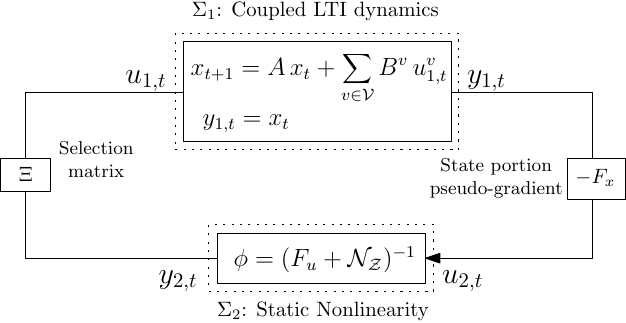}
\end{center}
\caption{Feedback interconnection between LTI system $\Sigma_1$ and the static nonlinearity $\phi(\cdot)$ in $\Sigma_2$. } \label{fig:FeedbackInt}
\end{figure}

Based on this definition of $\phi(\cdot)$, the mapping $\mathcal{S}$ corresponds to $\phi \circ (-F_x)$, and we can rewrite the parametrized equation in~\eqref{eq:VI} as 
\begin{align}
u = \phi(-F_x \x).
\end{align}
In turn, we can write the closed-loop system~\eqref{eq:ClosedLoop} as an LTI system 
\begin{align} \label{dfn:Sigma1} 
 \Sigma_1:\left\{
 \begin{array}{rl}x_{t+1} &= \displaystyle A\,x_{t} + \sum_{v\in \mc{V}} B^v \,u^v_{1, t}\\
y_{1, t} &= x_{t},\end{array}
\right.
\end{align}
in feedback with the static nonlinear map
\begin{align} \label{dfn:Sigma2}
\Sigma_2: y_{2,t} = \phi(u_{2,t}),
\end{align}
and connected through
\begin{align} \label{dfn:Interc}
u_{1,t} = \Xi~y_{2,t}, \quad 
u_{2,t} = -F_x~y_{1,t},
\end{align}
where $(u_{i,t},y_{i,t})$ are the input and output of system $\Sigma_i$ with \mbox{$i \in \{1,2\}$}. The matrix $\Xi$ given in~\eqref{eq:FeedbackLaw} selects the first entry of the  optimal predicted input sequence, and $F_x$ is the pseudo-gradient of the state cost over the horizon given in~\eqref{eq:Fx}. The whole interconnection~\eqref{dfn:Sigma1}--\eqref{dfn:Interc} is shown in Figure~\ref{fig:FeedbackInt}.

Building on this view of RHG in~\eqref{eq:ClosedLoop} as an interconnection of two dissipative systems, we proceed to establish our main stability result.
\begin{thm} \label{thm:StabCond}
Suppose that Assumption~\ref{ass:Global} holds, and that there exists $P \in \mathbb{S}_{\succ 0}^{n_x}$ and $\eta> 0$ such that 
%
\begin{align} \label{eq:StabCond}
\begin{bmatrix}
A^\top PA - P &  A^\top P\hat{B}\\
* &  \hat{B}^\top P\hat{B}
\end{bmatrix}  
+\eta \begin{bmatrix}0
& -\frac{F_x^\top}{2}\\
* & -\mu I \end{bmatrix} \preceq -\epsilon I,
\end{align}
for some $\epsilon>0$, where $\hat{B} = B\,\Xi$, $\,\mu$ is the strong monotonicity constant of the operator $F_u(\cdot)$ given in~\eqref{eq:Fu}, and $F_x$ as in~\eqref{eq:Fx}.
Then, the following conditions hold:
\begin{itemize}
\item[(i)] there exists a globally asymptotically stable equilibrium point, $\bar x \in \R^{n_x}$, of the closed-loop system~\eqref{eq:ClosedLoop}; 
\item[(ii)] the OCP in~\eqref{eq:MPCagent} is recursively feasible, $\forall\, \x\in \R^{n_x}$; and
\item[(iii)] the control inputs satisfy constraints for all times, $u_t\in \mc{Z}\; \forall t $.
\end{itemize}
\end{thm}
\begin{proof}
See Appendix~\ref{sec:ProofStabCond}.
\end{proof}
Closely related are the results in~\cite{hall2022receding, benenati2024probabilistic} which however both assume a potential game structure. The resulting generalized potential game, solved in a receding-horizon fashion, is equivalent to an economic MPC problem~\cite{faulwasser2018economic}, for which stability results without terminal ingredients have already been established~\cite[Theorem 4.1]{faulwasser2018economic}. In the non-potential game case considered here, this connection no longer holds, requiring the development of new stability results, as presented in Theorem~\ref{thm:StabCond} based on the monotonicity property of the pseudo-gradient in~\eqref{eq:PseudoGrad} and dissipativity of the LTI system~\eqref{eq:LTI}. Dissipativity generalizes both small-gain and passivity-based interconnection theorems. This analysis captures how the stable LTI system is affected by the control input generated by solving the non-cooperative game defined in the RHG feedback law in~\eqref{eq:FeedbackLaw}. Note that we do not require terminal ingredients to derive the stability certificate in~Theorem~\ref{thm:StabCond}.
Moreover, the certificate is easily verified as checking for the existence of $P$ and $ \eta$ can be posed as a semidefinite program and solved in Python using CVXPY~\cite{diamond2016cvxpy}.

The LMI in Theorem~\ref{thm:StabCond} gives a sufficient but not necessary condition for stability of~\eqref{eq:FeedbackLaw}. The results are conservative due to the analysis strategy which relies on dissipativity of interconnected systems and which treats the feedback term in $\Sigma_2$ as a perturbation to the coupled LTI dynamics $\Sigma_1$. 
\begin{rmk}
Generally, the strong monotonicity constant $\mu$ must be computed offline. In some cases, it can be deduced from the game's primitives~\cite{belgioioso2017convexity, scutari2014real} or properties of the Jacobian matrix $Jac \; F_u(\cdot)$~\cite{taji1993globally, bauschke2017convex}. In the case of a quadratic actuation cost, $Jac\; F_u(\cdot)$ is a constant matrix and $\mu$ can be computed using techniques for distributed eigenvalue estimation~\cite{gusrialdi2020data}.
\end{rmk}
Our stability analysis relies on an equilibrium-independent Lyapunov function and, in general, offers no characterization of the equilibrium point itself. However, an expression for $\bar x$ can be found if an additional terminal-state constraint is imposed.
\begin{prp}\label{prp:EquPoint}
Let Assumption~\ref{ass:Global} hold, and denote by $(u_s, x_s)$ the unique v-GNE of the steady-state game
\begin{align}\label{eq:EquilSteady}
v\in \mc{V}: \left\{
\begin{array}{r l}
\displaystyle \min_{u^v, x} & \; \ell^v(x, u^v, u^{-v}) \\ 
 \text{s.t} &  x =  A\,x + {\textstyle\sum_{v\in \mc{V}}} B^v\, u^v,\\ 
& u^v \in \mc{U}^v, (u^v,u^{-v})\in \mc{C}.\\
\end{array}  
\right.
\end{align} 
If $u_s \in \text{int}\, \mathcal{Z}$, then $(u_s, x_s)$ is the unique equilibrium point of the closed-loop system~\eqref{eq:ClosedLoop} for the RHG feedback law defined by the game in~\eqref{eq:MPCagent} with the additional terminal state constraint $x_K = x_s$.
\end{prp}
\begin{proof}
See Appendix~\ref{sec:ProofEquPoint}. 
\end{proof}
%
RHGs are a generalization of economic MPC (EMPC) to a game-theoretic setting. An RHG collapses into EMPC if there is only one agent, and we can use stability results for EMPC~\cite{hall2022receding} if the game admits a potential. For EMPC without terminal ingredients under dissipativity conditions, the equilibrium point $\bar{x}$ lies in a neighbourhood of the optimal steady-state operating point $x_s$~\cite[Thm 4.1]{faulwasser2018economic}, and $\bar x \to x_s$ as $K \to \infty$ under additional continuity assumptions~\cite[Lem 4.1]{faulwasser2018economic}. We conjecture a similar convergence behavior for the equilibrium of RHG policies~\eqref{eq:MPCagent} without terminal ingredients, but under a Lyapunov function decrease condition as in Theorem~\ref{thm:StabCond}. This conjecture is supported by our numerical simulations.
%


\section{Distributed stability condition}
%
The computational complexity of solving the LMI in~\eqref{eq:StabCond} increases significantly with the game size, rendering stability checks for large agent populations intractable. In this section, we explore a specialized setting that enables a distributed stability verification. Specifically, consider decoupled LTI dynamics of the form
\begin{align}\label{eq:DynDecomp}
x_{t+1}^v = A^v\,x^v_t + B^v\,u^v_t, \quad \forall v \in \mathcal{V},
\end{align}
and the stage costs decoupled in the state term, as follows: \begin{align}\label{eq:CostDecomp}
\ell^v(x^v_t,u^v_t, u^{-v}_t)=(x_t^v)^\top W^{v} x^v_t + (w^v)^\top x^v_t + \ell_u^v(u_t^v,u_t^{-v}).
\end{align}
The setup arises whenever agents control local mechanical systems, such as in robotics~\cite{gu2008differential}, and energy management applications~\cite{hall2022receding}. It allows one to solve for the LMI~\eqref{eq:StabCond} in a distributed manner.
\begin{cor}\label{cor:StabCondLocal} 
Let Assumption~\ref{ass:Global} hold, and let dynamics and stage costs be as in~\eqref{eq:DynDecomp}--\eqref{eq:CostDecomp}. 
Then, conditions (i)--(iii) of Theorem~\ref{thm:StabCond} hold if, for all agents $v \in \mc{V}$, there exists $P^v \in \mathbb{S}_{\succ 0}^{n_x^v}$ and $\eta^v > 0$ 
such that the following local LMI is verified:
\begin{align} \label{eq:StabCondLocal}
&\begin{bmatrix}
(A^v)^\top P^vA^v - P^v &  (A^v)^\top P^v\hat{B}^v\\
* &  (\hat{B}^v)^\top P^v \hat{B}^v\\
\end{bmatrix}+  \eta^v \begin{bmatrix}0  & -\frac{(F_x^v)^\top }{2}\\
* & -\mu I \end{bmatrix} \preceq - \epsilon I,
\end{align}
%
%
for some $\epsilon>0$, where $\hat{B}^v = B^v\,\Xi^v$, $\mu$ is the strong monotonicity constant of $F_u(\cdot)$ given in~\eqref{eq:FuDist}, and $F^v_x$ is defined in~\eqref{eq:FxDist}.
\end{cor}
\begin{proof}
The proof is given in~Appendix~\ref{sec:DistrStabCheck}.
\end{proof}
%
%
Corollary~\ref{cor:StabCondLocal} provides a method for the agents to verify closed-loop stability of their local dynamics under the RHG feedback law in~\eqref{eq:FeedbackLaw}. Practically, to verify the LMI condition in~\eqref{eq:StabCondLocal}, each agent needs to solve a convex optimization problem locally. For this purpose, we assume that agents have knowledge of $\mu$, the strong monotonicity constant of the pseudo-gradient $F$.

After presenting the distributed stability check, the question arises as to what happens if the LMI in~\eqref{eq:StabCondLocal} is infeasible and how agents can modify their local subsystems to ensure stability. To gain such insights, we analyze the 1-dimensional case in the following corollary.  

\begin{cor}\label{cor:1DstabInequ}
For all $v \in \mc{V}$, let $A^v, B^v \in \mathbb{R}$, and assume there exist $\eta^v>\frac{(B^v)^2}{\mu}> 0$ such that the following condition holds:
\begin{equation}\label{eq:1Dcond}
\frac{\eta^v\mu (A^v)^2}{\eta^v\mu -   (B^v)^2} + \frac{\eta^v (B^v)^2 (W^v)^2}{\mu}   \sum_{k=0}^{K-2}((k+1)(A^v)^k)^2 < 1.
\end{equation}
%

Then, conditions (i)--(iii) of Theorem~\ref{thm:StabCond} are satisfied. 
\end{cor}

\begin{proof} 
Proof in Appendix~\ref{sec:Proof1DIneq}.
\end{proof}

\begin{figure}
     \centering
     \begin{subfigure}[b]{0.21\textwidth}
         \centering
         \includegraphics[width=\textwidth]{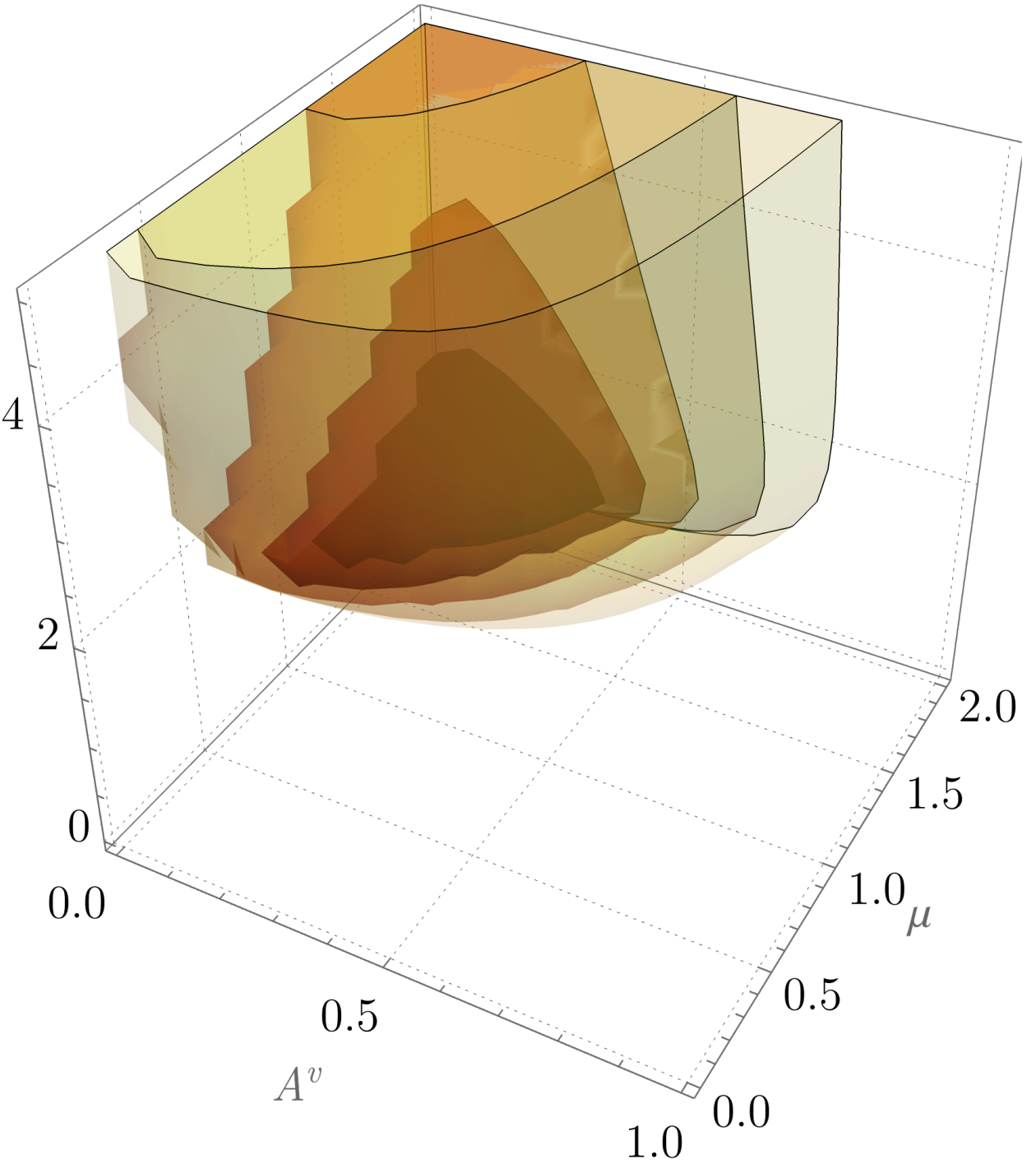}
         \caption{3D feasibility region of~Corollary~\ref{cor:1DstabInequ} for $\mu, A^v, \eta^v$.}
         \label{fig:Feasibility3D}
     \end{subfigure}\hspace{0.5cm}%
     \begin{subfigure}[b]{0.24\textwidth}
         \centering
         \includegraphics[width=\textwidth]{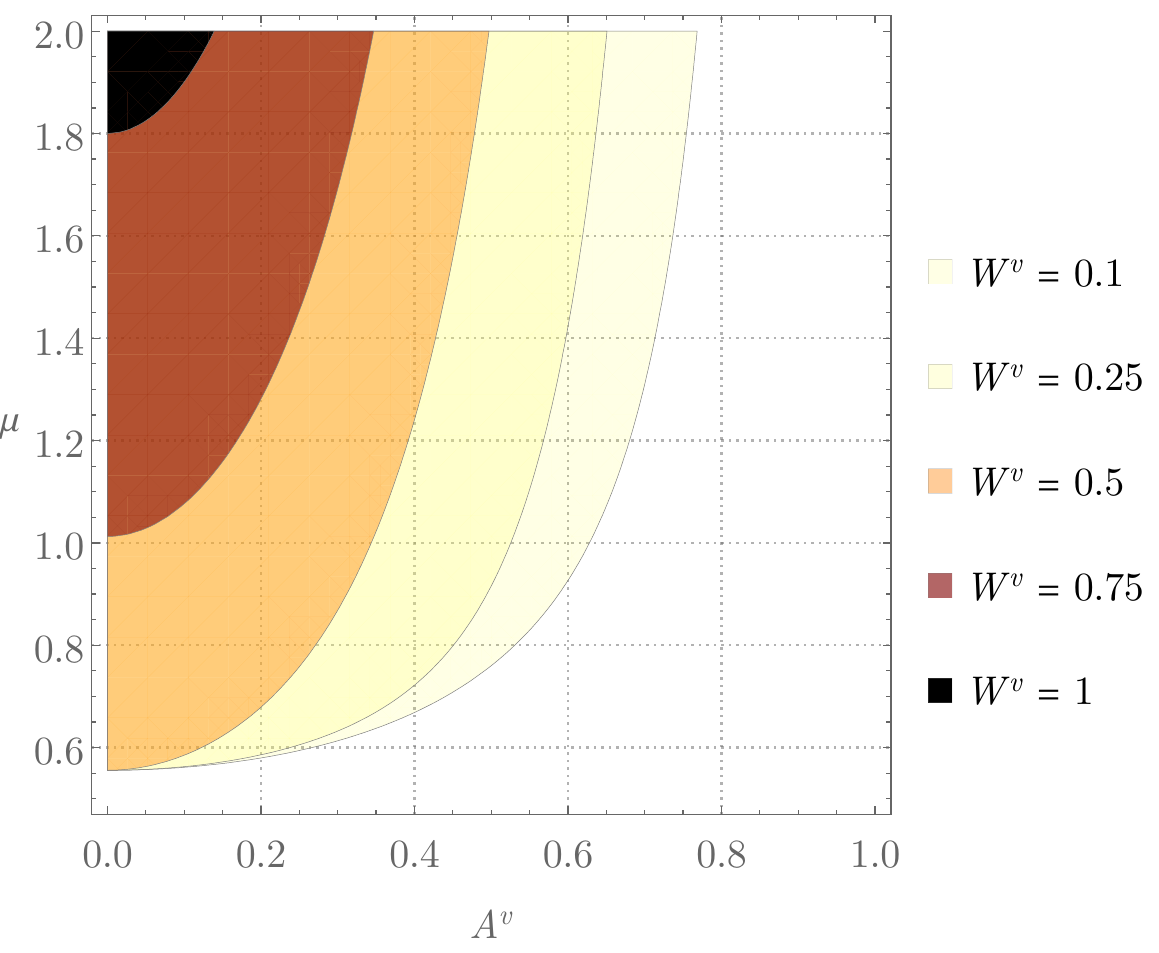}
         \caption{2D cross-section of (a) for \mbox{$\eta^v= 1.80$}.}
         \label{fig:Feasibility2D}
     \end{subfigure}
     \vspace{17 pt}
     \caption{Feasibility regions of Corollary~\ref{cor:1DstabInequ} as a function of $A^v$, $\eta^v$, $\mu$, and $W^v$ for fixed $K=10$ and $B^v = 1$. The plots were generated using the RegionPlot function of Mathematica.} \label{fig:Feasibility}
\end{figure}

The feasibility regions of the inequality~\eqref{eq:1Dcond} are shown in Figure~\ref{fig:Feasibility} as a function of $A^v$, $\mu$, and $W^v$, as well as the optimization variable $\eta^v$. Note that decreasing $W^v$ and $A^v$ in~\eqref{eq:1Dcond} makes the left-hand side smaller, which is an observation confirmed by Figure~\ref{fig:Feasibility}.
%
%
This one-dimensional analysis provides valuable insights that we use to develop general tuning guidelines that favor the feasibility of the local LMI stability condition in~\eqref{eq:StabCondLocal}. Specifically:
\begin{enumerate}
\item[(i)] Decrease the magnitude of state weighting, i.e., reduce $\rho(W^v)$.
\item[(ii)] Increase stability of the local dynamics, i.e., decrease $\rho(A^v)$.\end{enumerate}
%
We will demonstrate how to apply these tuning guidelines through an illustrative example in the upcoming section.
\section{Numerical case studies}
\subsection{Illustrative example: RHG can destabilize stable systems}
\label{ss:IE}
Consider two agents with local dynamics of the form~\eqref{eq:DynDecomp},
with 
\begin{align*}
A^1 = \left[\begin{smallmatrix}0.6& 0.3\\ 0.3& 0.7 \end{smallmatrix}\right], \; B^1 = \left[\begin{smallmatrix}10 & 5.5\\ 11&  4\\\end{smallmatrix}\right],\;
A^2 = \left[\begin{smallmatrix}0.6&  0.1\\
0.8& 0.1\end{smallmatrix}\right], \; B^2 = \left[\begin{smallmatrix}13&19\\6.5& 10\\\end{smallmatrix}\right],
\end{align*}
yielding $\rho(A^1)=0.954$ and $\rho(A^2) = 0.727$. Both agents have local state cost and an input coupling cost of the following form:
\begin{align}
\ell(x_t^v, u_t^v, u_t^{-v}) = (x_t^v)^\top W^v (x_t^v)+  \Big(\sum_{j \in \mc{V}}R^{v} u^j \Big)^\top u^v,
\end{align}
where the parameters of the input cost are $R^1 = \textrm{blkdiag}(10,0.01)$ and $R^2 = \textrm{blkdiag}(0.01,20)$. Cost functions with this structure characterize aggregative games~\cite{belgioioso2017convexity}, and appear in many resource allocation problems.

%

We simulate the resulting closed-loop system~\eqref{eq:FeedbackLaw} using a prediction horizon of $K=10$, and different $W^v$ matrices. Our findings are twofold. Firstly, we observe that the agents can destabilize each other by using RHG, even though the local dynamics are stable, as shown in Figure~\ref{fig:ToyA}. This underscores the importance of our stability analyses and certificates. Secondly, we observe in Figure~\ref{fig:ToyC} that with a decreasing $\rho(W^v)$ the closed-loop dynamics become stable, in line with our tuning guidelines. Additionally, we observe that:

\begin{figure}
\centering
 \begin{subfigure}[a]{0.5\textwidth}
 \centering
\includegraphics[scale=0.46]{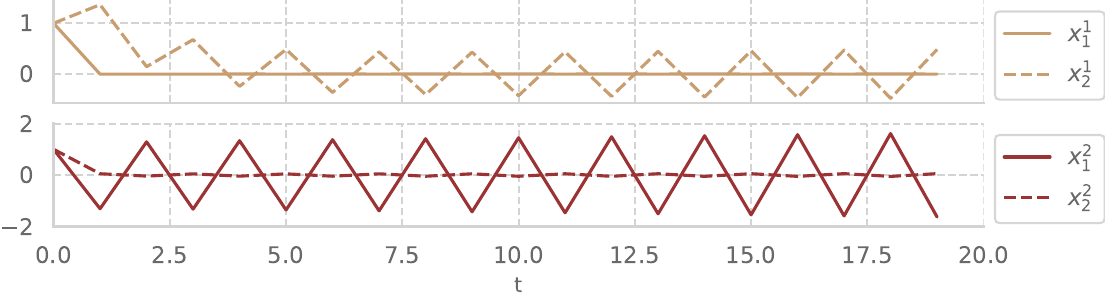}
 \caption{$W^1 = \text{diag}(20,0.05), W^2 = \text{diag}(0.05,20)$, $\mu= 0.498$.}
 \vspace{17 pt}
\label{fig:ToyA}
 \end{subfigure}
 \begin{subfigure}[b]{0.5\textwidth}
 \centering
\includegraphics[scale=0.46]{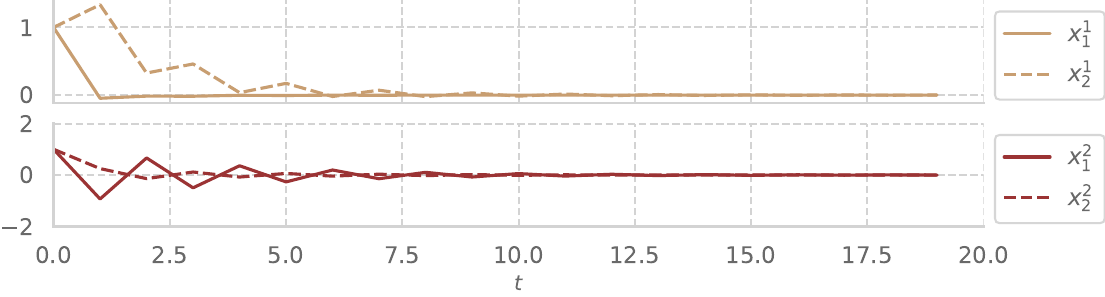}
\caption{$W^1 = \text{diag}(1,0.05), W^2 = \text{diag}(0.05,1)$, $\mu= 0.125$.}
\label{fig:ToyC}
\end{subfigure}
\vspace{12 pt}
\caption{Evolution of the closed-loop system in Section~\ref{ss:IE}, for different choices of state weights $W^v$ in the RHG feedback law. \label{fig:Toy}}
\end{figure}

\begin{itemize}
\item In contrast to classical MPC, we observed in simulations that increasing the horizon length $K$ does not have a stabilizing effect for the closed-loop system under the RHG feedback law~\eqref{eq:FeedbackLaw}. 
\item The feasibility of the LMI is negatively influenced not only by the magnitude of state and input weights, i.e., $\rho(R^v), \; \rho(W^v)$, but also by their asymmetry across different agents. 
\item The stable closed loop in Figures~\ref{fig:ToyC} stabilizes to the corresponding steady-state v-GNE~\eqref{eq:EquilSteady}, i.e., as $t \to \infty$ then $x^v\to \bar{x}^v = x_s^v = [0,~0]^\top$. This supports our conjecture that agents find the operating point $x_s$ without terminal constraints. 

\end{itemize}


Finally, our numerical studies highlight the conservative nature of the stability certificates. Specifically, we have observed instances where, despite violations of the stability certificates, the closed-loop systems remain stable. Moreover, while our analysis requires strong monotonicity of the pseudo-gradient, as postulated in Assumption~\ref{ass:Global}(iii), the simulations reveal that stable closed-loop systems can be achieved even in scenarios with non-monotone pseudo-gradients.

\subsection{Battery-charging game} \label{sec:BCG}
In this section, we deploy RHG on a simplified version of the battery charging game presented in~\cite{atzeni2012demand,hall2022receding}.
 \begin{figure}[b]
 \centering
\includegraphics[width=0.9\columnwidth]{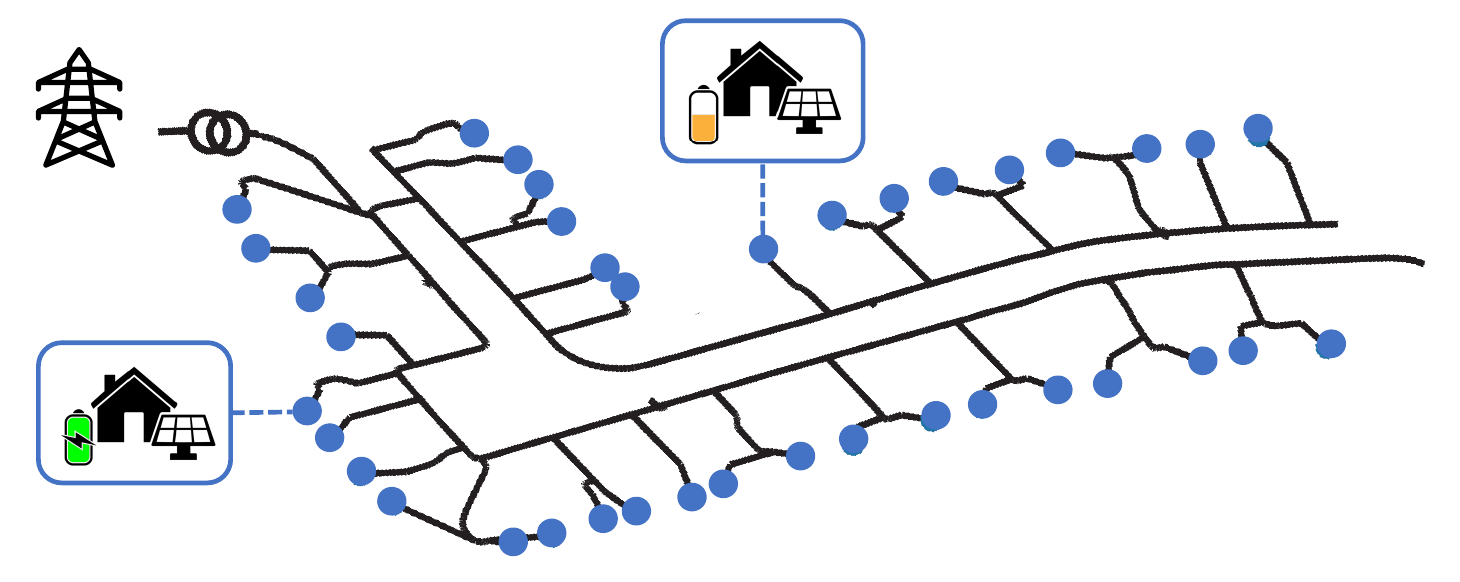}
 \caption{Schematic of households owning a private battery and being connected to the power grid through a point of common coupling.}
\label{fig:BatterySchem}
 \end{figure}
\subsubsection{Model}
We consider a distribution grid composed of $M$ active consumers $v \in \mc{V} = \{1,\dots, M\}$ connected to the main transmission grid via a point of common coupling, as is shown in Figure~\ref{fig:BatterySchem}. 
Every consumer  $v\in \mc{V}$ owns an energy storage device, or battery, with a controllable input $u_t^v$ modeling charging if $u_t^v > 0$ and discharging if $u_t^v < 0$.
Each consumer has an inflexible energy demand of $d^v_t$ units of power, at time $t$, which they must fulfill by buying energy from the main grid $l_t^v \in \bb{R}_{\geq 0 }$ and/or discharging their local battery $u_t^v$. This requirement translates in the local energy-balance constraint
\begin{align} 
 l_t^v = u_t^v + d_t^v , \quad \forall v \in \mc{V} \label{eq:LoadGen}.
\end{align}
%
The power supplied to individual consumers is limited by
\begin{equation} \label{eq:LoadLim}
0 \leq l_t^v \leq l_{\text{max}}^v, \quad \forall v \in \mc{V},
\end{equation}
where $l_{\text{max}}^v$ is the maximum power a consumer can absorb from the main grid over the sampling period. Moreover, the charging/discharging control input $u_t^v$ is constrained by the charging rate:
\begin{align}  \label{eq:ChargeCon}
u_{\text{min}}^v\leq  u_t^v \leq u_{\text{max}}^v,
\end{align}
where $ u_{\text{max}}^v\geq 0 \geq u_{\text{min}}^v$ are the upper and lower charging limits. Thus, the local input constraint set~\eqref{eq:LocalConstr} is the following:
\begin{align*}  
\mc{U}_t^v= \{u^v_t ~|~ 0 \leq u_t^v+  d_t^v \leq l_{\text{max}}^v, ~ u_{\text{min}}^v\leq  u_t^v \leq u_{\text{max}}^v\}.
\end{align*}
Power line and transformer constraints at the point of common coupling limit the aggregate load of all consumers, as follows:
\begin{equation} \label{eq:AggrLoadLim}
0 \leq \sum_{v\in \mc{V}}l_t^v \leq L_{\text{max}},
\end{equation}
where $L_{\text{max}}>0$ is the grid capacity. This constraint constitutes a coupling constraint, as in~\eqref{eq:CouplingConstr}. The corresponding constraint set is
\begin{align*} 
\mc{C}_t= \left\{(u^v_t,u^{-v}_t) ~\Big|~ 0 \leq \sum_{v\in \mc{V}} u_t^v + d_t^v \leq L_{\text{max}}\right\}.
\end{align*}
Each battery has a scalar state-of-charge (SoC) $x_t^v$ evolving as
\begin{align} \label{eq:BatteryDyn}
x_{t+1}^v = A^v x_t^v + B^v u_t^v,  \quad \forall v \in \mc{V},
\end{align}
where $A^v, B^v \!\in\! \left[0, 1\right]$ model leakage rate and charging efficiency. 

We assume that each consumer $v \in \mc{V}$ is self-interested and aims to minimize its electricity bill as well as the operational cost of its battery, yielding a local cost function of the form
\begin{align*} 
\ell_t^v(x^v_t,u^v_t,u_t^{-v}) =\,  \underbrace{\left(\gamma_{1,t}^v \sum_{j\in \mc{V}}l_t^j+ \gamma_{2,t}^v \right) \,l_t^v}_{\text{energy cost}} +\underbrace{\gamma_{3,t}^v \, (x^v_t-x^{\text{ref}})^2}_{\text{battery usage}},
\end{align*}
where $l_t^v = u_t^v + d_t^v$ as defined in~\eqref{eq:LoadGen}. This stage cost follows the general structure in~\eqref{eq:StageCost}. The constants  $\gamma_{1,t}^v, \gamma_{2,t}^v>0$ represent different price rates that consumers previously negotiated with suppliers. 
The second term, with $\gamma_{3,t}^v>0$, reflects agent $v$'s aim to maintain its battery charge state near the reference $x^{\text{ref}}$ to minimize degradation.

\subsubsection{Implementation}
\begin{figure}
 \centering
\includegraphics[scale=0.38]{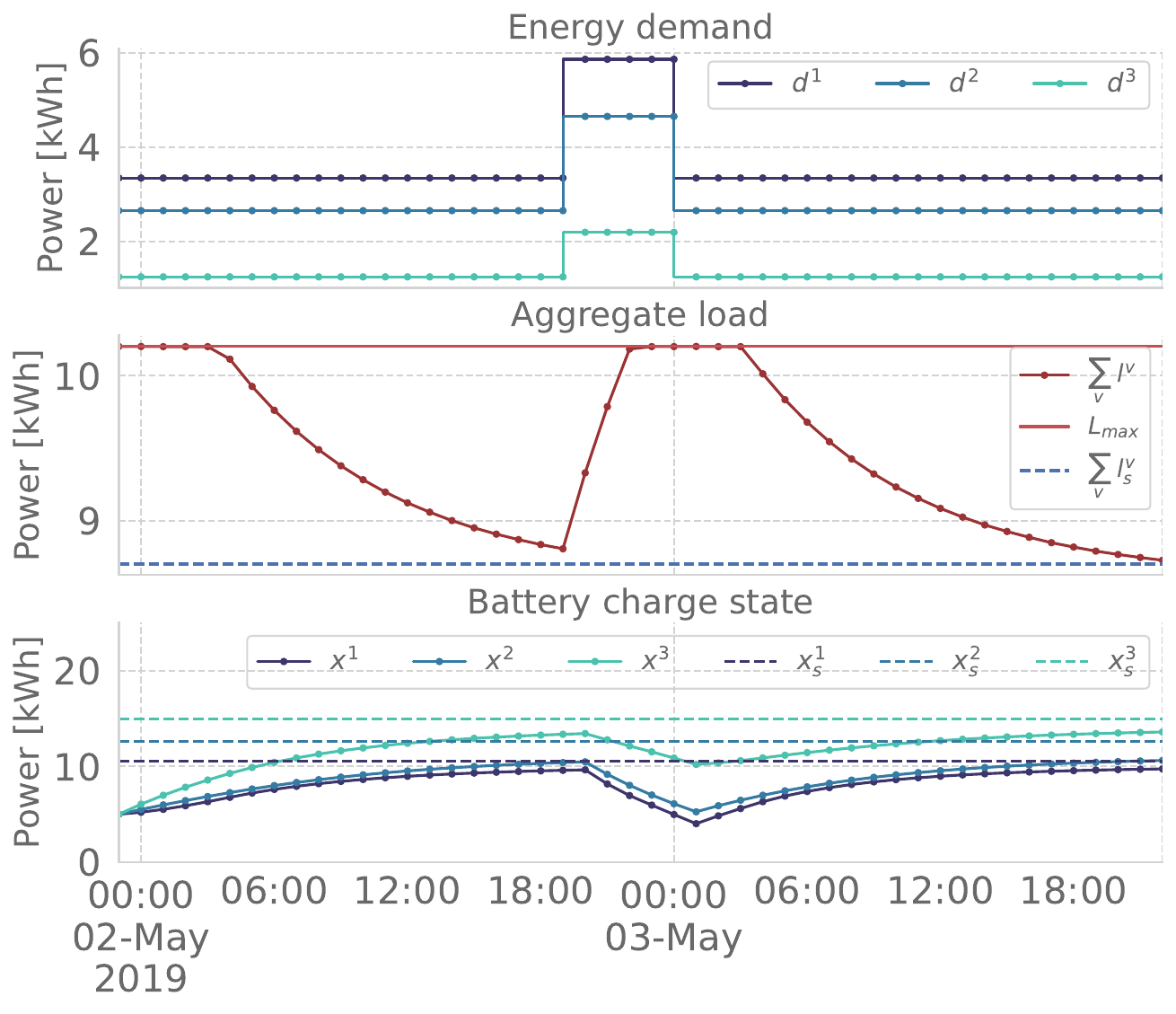}
 \caption{Evolution of the energy demand (top), aggregate load (center), and battery state of charge (bottom) for the 3-consumer charging game in Section~\ref{sec:BCG} under the RHG feedback law.}
\label{fig:BatterySim}
\end{figure}

We consider a battery charging game between 3 heterogeneous consumers owning similar storage devices, i.e., a lithium-ion battery with SoC dynamics as in (\ref{eq:BatteryDyn}) and parameters $A^v \in [0.955, 0.98]$ (corresponding to a leakage rate of about 0.8 over 24 hours) and $B^v \in [0.7,0.9]$, as in~\cite{atzeni2012demand}, $u_{\text{max}}^v = -u_{\text{min}}^v = 7$~kWh, and $x^{\text{ref}}\in [15,20]$ kWh. The price rates are \mbox{$\gamma^v_1 \in [0.03,0.05]$~\textdollar/kWh} for the coupling price, $\gamma^v_2 \in [0.01,0.2]$ \textdollar/kWh, for the base energy price, and $\gamma_{3}^v \in [0.01, 0.02]$ for battery usage weight. For the demand profiles $d^v_t$, we used real data from single-family homes situated in the state of New York collected on the 2nd of May 2019 and averaged over 48 hours~\cite{pecanstreet2022dataport}. Additionally, we apply an energy demand shock which doubles the base demand $d_t^v$ of all agents between 9 pm and 1 am. After successfully verifying closed-loop stability using the local LMI-based certificates in Corollary~\ref{cor:StabCondLocal}, we simulate this battery charging game under RHG control over 48 hours, and illustrate the results in Figure~\ref{fig:BatterySim}. All simulations are implemented in Python, and the \mbox{v-GNE} is computed using a smoothed \mbox{Fischer-Burmeister} method~\cite{liaomcpherson2019regularized}.

\subsubsection{Discussion}
The simulation study demonstrates three things: (i) The distributed stability check in Corollary~\ref{cor:StabCondLocal} effectively guarantees closed-loop stability, (ii) the consumers stabilize to the steady-state v-GNE~\eqref{eq:EquilSteady}, as conjectured; (iii) the aggregate load constraint~\eqref{eq:AggrLoadLim} can be enforced despite unforeseen disturbances. Before the demand shock, all consumers charge their batteries, so that the SoC converges to a close neighborhood of the steady-state \mbox{v-GNE} $x_s^v$~\eqref{eq:EquilSteady}, as shown in Fig.~\ref{fig:BatterySim} (bottom). 
During the demand shock, the agents buy from the grid and discharge their batteries to serve their increased demand. Note that the aggregate load constraint is successfully enforced, albeit agents have no preview of the shock. After the demand shock, states and inputs re-stabilize to their respective steady states.

\section{Conclusion}

This paper presents LMI-based stability certificates for game-theoretic MPC, specifically for strongly monotone games with coupling in the cost and the constraints played in receding horizon. Our certificates can be numerically verified to show global asymptotic stability. Moreover, if the agents have decoupled dynamics, the certificates can be evaluated locally, therefore offering a simple path towards verifying asymptotic stability for games with a large number of agents. We demonstrate the usefulness of RHG and the effectiveness of our certificates in a battery charging game. In future work, we will focus on extending the stability analysis beyond open-loop stable systems by exploring nonlinear storage functions and developing suitable terminal ingredients.

\bibliography{PhDLiterature}{}

\appendix
\section{Appendix}
\subsection{Proof of Theorem~\ref{thm:StabCond}} 
\label{sec:ProofStabCond}
We start by writing the pseudo-gradient~\eqref{eq:PseudoGrad} more explicitly as
\begin{align}\label{eq:PseudoGradDetailed}
F(u,\x) = F_u(u) + F_x \x,
\end{align}
where the component mappings are defined in \eqref{eq:Fu} and \eqref{eq:Fx}.
%
The proof of Theorem~\ref{thm:StabCond} consists of four main parts. 1) Showing existence of the equilibrium point $\bar x$, 2) Deriving dissipativity inequalities for $\Sigma_1$ and $\Sigma_2$, 3) Interconnecting the two subsystems, 4) Invoking~\cite[Th. 13.2]{haddad2008stability} to verify that the sum of storage functions of the two subsystems fulfills the conditions for a Lyapunov function.

1) We begin by showing that the set of equilibrium points for~\eqref{eq:ClosedLoop}, $\mc{E} = \{\bar x~|~\bar x = A\bar x + B\kappa(\bar x)\}$, is non-empty by using a fixed point theorem~\cite{morris1975schauder}. Rearranging the definition of $\mc{E}$, we define the mapping
$z(x) = (I-A)^{-1} B\kappa(x)$ which: (i) is single-valued (as $\rho(A)<1$ by Assumption~\ref{ass:Global}~(i)), (ii) has the same set of fixed-points as the closed-loop dynamics~\eqref{eq:ClosedLoop}, (iii) is continuous by Proposition~\ref{prp:PhiProperties} and (iv) maps onto the set $\{z~|~ (I-A) z = Bu,~~u\in \mc{Z}\}$ which is compact as $\mc{Z}$ (defined in~\eqref{eq:GlobalDecision}) is compact under Assumption~\ref{ass:Global}~(ii). Hence, we can apply the Schauder-Tychonoff fixed-point theorem~\cite{morris1975schauder} to conclude existence of at least one fixed point $\bar x \in \mc{E}$ of $z(x)$ which is also an equilibrium point of~\eqref{eq:ClosedLoop}.

2) Next, we derive dissipation inequalities with storage functions $V_1 $ and $V_2$ for each subsystem $\Sigma_1$~\eqref{dfn:Sigma1} and $\Sigma_2$~\eqref{dfn:Sigma2} separately. We define $\Delta x_t = (x_t - \bar x )$, $\Delta y_{t,i} = (y_{t,i} - \bar  y_i)$, and $\Delta u_{t,i} = (u_{t,i} - \bar  u_i)  ,~ i \in \{1,2\}$. As $\Sigma_1$ is a discrete-time LTI system, we choose a quadratic storage function family as $V^1_{\bar x}(x_t) = (\Delta x_t)^\top P~(\Delta x_t)$, for some positive definite $P$. Evaluating the evolution of $V^1_{\bar x}$ along trajectories gives 
%
%
%
%
\begin{multline} \label{eq:DissSigma1}
V^1_{\bar x}(x_{t+1}) - V^1_{\bar x}(x_{t}) = \\\begin{bmatrix}
\Delta y_{1,t}\\
\Delta u_{1,t} 
\end{bmatrix}^\top
\begin{bmatrix}
A^\top PA -  P &  A^\top PB\\
* &  B^\top PB\\
\end{bmatrix}
\begin{bmatrix}
\Delta y_{1,t}\\
\Delta u_{1,t}
\end{bmatrix}
\end{multline}
%
%
By Proposition~\ref{prp:PhiProperties}, the static nonlinearity $\phi(\cdot)$ in~\eqref{dfn:Sigma2} is $\mu$-cocoercive 
and thus obeys the following inequality where $\eta> 0$:
\begin{align}\label{eq:Diss1}
V^2_{\bar x} = 0 \leq&  \, \eta \begin{bmatrix}
\Delta y_{t,2}\\
\Delta u_{t,2}
\end{bmatrix}^\top\ \begin{bmatrix}- \mu I & \frac{1}{2} I\\
* & 0\end{bmatrix} \begin{bmatrix}
\Delta y_{t,2} \\
\Delta u_{t,2}
\end{bmatrix}.
\end{align}

\noindent 3) To analyze the interconnection of the two systems we recall that
\begin{align}\label{eq:Intercon1}
\Delta u_{1,t} = \Xi~ \Delta y_{2,t}\\ \label{eq:Intercon2}
\Delta u_{2,t} = -F_x~ \Delta y_{1,t}.
\end{align}
By substituting~\eqref{eq:Intercon1} into~\eqref{eq:DissSigma1} and~\eqref{eq:Intercon2} into~\eqref{eq:Diss1}, we can write both inequalities in terms of the outputs $\Delta y_{1,t}$ and $\Delta y_{2,t}$, and add them together which gives
{\begin{small}
\begin{align*}
&V_{\bar x}(x_{t+1}) - V_{\bar x}(x_{t})= \, V^1_{\bar x}(x_{t+1}) - V^1_{\bar x}(x_{t}) + 0 \\
&=\begin{bmatrix}
\Delta y_{1,t}\\\nonumber
 \Delta y_{2,t}
\end{bmatrix}^\top \left( \begin{bmatrix}
A^\top PA -  P &  A^\top P\hat{B}\\
* &  \hat{B}^\top P\hat{B}\\
\end{bmatrix} + \eta \begin{bmatrix}0  & -\frac{1}{2}F_x^\top\\
* & - \mu I \end{bmatrix}\right)
\begin{bmatrix}
\Delta y_{1,t}\\
 \Delta y_{2,t}
\end{bmatrix},
\end{align*}\end{small}}
where $\hat{B} = B~\Xi$. Thus, if 
\begin{align}\label{eq:Vdecrease}
\begin{bmatrix}
A^\top PA - P &  A^\top P\hat{B}\\
* &  \hat{B}^\top P\hat{B}\\
\end{bmatrix} + \eta \begin{bmatrix}0  & -\frac{1}{2}F_x^\top\\
* & -\mu I \end{bmatrix} \preceq -\epsilon\, I
\end{align}
for some $\epsilon >0$, then $V_{\bar x}(x_{t+1}) - V_{\bar x}(x_{t})< 0$. 

4) To show closed-loop stability of the interconnected system, we apply the discrete-time Lyapunov theorem~\cite[Thrm. 13.2]{haddad2008stability}. To invoke this result, the following conditions must be satisfied: (i)~$V_{\bar{x}}(\bar{x})=0$, (ii)~$V_{\bar{x}}(x)> 0, \;\forall x\in \R^{n_x}\backslash \{\bar{x}\}$, (iii)~$V_{\bar{x}}(x)\to~\infty$ as $\|x\|\to \infty$, (iv)~$V_{\bar{x}}(f(x))-V_{\bar{x}}(x)< ~0,$ \mbox{$\forall x\in \R^{n_x}$}.
Requirements (i)-(iii) hold as the candidate function is defined as $V_{\bar{x}}(x_t) = \Delta x^\top P \Delta x$ with $P \in \mc{S}_{\succ 0}^{n_x}$, and (iv) is satisfied as we impose the decrease in~\eqref{eq:Vdecrease}. Therefore, all conditions of~\cite[Thrm. 13.2]{haddad2008stability} are fulfilled and the equilibrium point $\bar{x}$ is globally asymptotically stable for $\epsilon > 0$ and stable if $\epsilon = 0$. 
Input constraint satisfaction is always ensured by the solution of the OCP in~\eqref{eq:MPCagent} and the problem is recursively feasible $\forall\, \x \in \R^{n_x}$ as we have no state constraints.
%
\subsection{Proof of Proposition~\ref{prp:EquPoint}}
\label{sec:ProofEquPoint}
We want to prove that the steady-state v-GNE $x_s$ in~\eqref{eq:EquilSteady} is the unique equilibrium point of the closed-loop system~\eqref{eq:ClosedLoop}, i.e., $x_s = A\,x_s + B\, \kappa_s(x_s)$,
where $\kappa_s(x)$ is the associated steady-state feedback law of the following RHG scheme
\begin{subequations} \label{eq:RHGTermConst}
\begin{align}
\min_{x,u^v}&\sum_{k=0}^{K-1}x_k^\top W^v x_k +(w^v)^\top x_k+ \ell_u^v(u^v_k,  u_s^{-v})\\ 
\textrm{s.t.}&\quad x_{k+1}=  A\, x_k + B^v\, u_k^v + B^{-v} u_s^{-v} , \quad k \in \bb{Z}_{K}\\
&\quad (u_k^v, u_s^{-v}) \in \mc{Z}, \label{eq:incs}\hspace*{9.3em} k \in \bb{Z}_{K}\\
&\quad x_0 = x_s, x_K = x_s,
\end{align}
\end{subequations}
which is the same as~\eqref{eq:MPCagent} but with an additional terminal state constraint $x_K = x_s$.

The proof consists of two main parts: 1) We derive the KKT conditions of the steady-state game~\eqref{eq:EquilSteady} and evaluate them at the proposed equilibrium point $(x_s, u_s)$. 2) We derive the KKT conditions of~\eqref{eq:RHGTermConst} and show  that $u^v_k= u^v_s,\, \lambda_k  = \lambda_s , \, \forall\,k \in  \bb{Z}_{K}$ and $x_k = x_s, $\mbox{$\forall\,k \in \bb{Z}_{K+1}$} is the unique v-GNE trajectory satisfying them and thus $\kappa_s(x_s) = u_s$. Since $u_s \in \text{int}\, \mathcal{Z}$ by assumption, we can neglect the inequality constraints in~\eqref{eq:incs} moving forward. 

1) We begin by noting that under Assumption~\ref{ass:Global}(iii), the pseudo-gradient mapping of~\eqref{eq:EquilSteady} is strongly monotone and thus the solution of the steady-state game~\eqref{eq:EquilSteady} is unique. We continue by formulating the following steady-state problem for each agent:
\begin{equation}v\in \mc{V}: \quad \left\{
\begin{array}{r l}
\displaystyle \min_{x,u^v}&\quad x^\top W^v x \,+\,(w^v)^\top x+\, \ell_u^v(u^v, u_s^{-v})\\ \label{eq:SteadyPerAgent}
\textrm{s.t.}&\quad x=  A\, x + B^v\, u^v + B^{-v} u_s^{-v},
\end{array}
\right.
\end{equation}
where $B^{-v} = [B^1, \dots, B^{v-1}, B^{v+1},\dots B^{M}]$. Strong duality of~\eqref{eq:SteadyPerAgent} holds as the objective function is convex in $x$ and $ u^v$, by Assumption~\ref{ass:Global}, and Slater's condition holds as the equality constraint is affine and feasible~\cite[Thm. 2.6.4]{goodwin2005constrained}. Thus, the KKT systems
\begin{subequations}
\begin{align}\label{eq:KKT1}
&2W^vx_s + w^v + (I-A)^\top \lambda_s = 0,\\ \label{eq:KKT2}
& \nabla_{u^v}  \ell_u^v(  u_s^v,  u_s^{-v}) -(B^v)^\top \lambda_s = 0,\\ \label{eq:KKT3}
&  x_s- A\, x_s + B^v\,  u_s^v + B^{-v} u_s^{-v} = 0.
 \end{align}
\end{subequations}  
for all $v \in \mathcal{V}$, are necessary and sufficient for $x_s, u^v_s, \lambda_s$ to be a v-GNE of the steady-state game~\eqref{eq:SteadyPerAgent}. 

2) Next, we derive the Lagrangian of~\eqref{eq:RHGTermConst} as
\begin{align*}
L^v(x,u^v,\lambda, \mu, \nu )= &\sum_{k=0}^{K-1}x_k^\top W^v x_k +(w^v)^\top x_k + \ell^v_u(u^v_k, u_s^{-v})   \\
&+\sum_{k=0}^{K-1} \lambda_k^\top (x_{k+1}- Ax_k + B^v u_k^v + B^{-v} u_s^{-v}) \\
&+ \mu^\top (x_s -x_K) + \nu^\top (x_0- x_s) ,
\end{align*}
and the resulting KKT conditions are
%
\begin{subequations}
\begin{align}
&\nabla_{x_K} L^v = \lambda_{K-1}-\mu = 0,\\  
&\nabla_{x_{k}} L^v  = 2\,W^v x_k+ w^v + \lambda_{k-1} - A^\top \lambda_{k}=0, \\
&\quad \quad  \forall k \in  \left\{\begin{smallmatrix}1,\dots, K-1\end{smallmatrix}\right\},\\
&\nabla_{x_{0}} L^v = 2\,W^v x_0 + w^v - A^\top \lambda_{0} +\nu=0,\\
&\nabla_{u_{k}^v} L^v = \nabla_{u^v_k} \ell_u^v(u^v_{k},u_s^{-v}) - (B^v)^\top \lambda_{k}=0, \, k \in \bb{Z}_{K},\\
&\nabla_{\lambda_{k}} L^v =  x_{k+1}- Ax_k + B^v u_k^v + B^{-v} u_s^{-v} = 0, \,  k \in \bb{Z}_{K}.
\end{align} 
\end{subequations}
By~\cite[Thm 2.5.6]{goodwin2005constrained}, these conditions, for all $v \in \mc V$, are necessary and sufficient for $\{x_k\}_{k=0}^K$, $\{u^v_k\}_{k=0}^{K-1}$ and $\{\lambda_k\}_{k=0}^{K-1}$ to be the unique v-GNE of~\eqref{eq:RHGTermConst}. Substituting $x_k = x_s, u_k = u_s$ and $\lambda_k = \lambda_s, \forall k \in \bb{Z}_{K}$, as well as $x_K=x_s, \mu = \nu = \lambda_s$ into the KKT conditions yields
%
\begin{subequations}
\begin{align}
& \lambda_{K-1} = \mu =  \lambda_s,\\
 \text{By }~\eqref{eq:KKT1}:\,  & 2\,W^v x_s \,+\,w^v+ (I-A)^\top  \lambda_s =0, \\ \nonumber
 &\quad \quad  \forall k \in  \left\{\begin{smallmatrix}1,\dots, K-1\end{smallmatrix}\right\},\\
\nonumber
 \text{By }~\eqref{eq:KKT1}: \, & 2\,W^v x_0+w^v - A^\top \lambda_{0} +\nu=0, \; \lambda_0 = \nu = \lambda_s,  \\ & \nonumber 2\,W^v  x_s +w^v -A^\top  \lambda_s + \lambda_s =0 \\ 
  &\Rightarrow  2\,W^v  x_s \; + w^v + (I-A)^\top  \lambda_s =0, \\
 \text{By }~\eqref{eq:KKT2}:\, & \nabla_{u^v_k} \ell_u^v(  u_s^v,u_s^{-v}) - (B^v)^\top \lambda_s = 0 , \; k \in \bb{Z}_{K},\\
 \text{By }~\eqref{eq:KKT3}:\, & (x_s- Ax_s + B^v u_s^v + B^{-v} u_s^{-v}) = 0, \;  k \in \bb{Z}_{K}.
\end{align}
\end{subequations}
Thus, the solution trajectories $u^v_k= u^v_s,\, \lambda_k  = \lambda_s , \, \forall\,k \in  \bb{Z}_{K}$ and $x_k = x_s, $\mbox{$\forall\,k \in \bb{Z}_{K+1}$} satisfy the KKT conditions for every agent. It follows that $\kappa_s(x_s) = u_s$ and  $(x_s, u_s)$ is the unique equilibrium of~\eqref{eq:ClosedLoop} with the terminal state constraint $x_K = x_s$.

\subsection{Proof of Corollary~\ref{cor:StabCondLocal} }
\label{sec:DistrStabCheck}

For the decoupled dynamics in~\eqref{eq:DynDecomp} and decoupled state cost~\eqref{eq:CostDecomp}, we define the mappings $F_u(u)$ and matrices $F_x^v, \; \forall v\in \mc{V}$,  as follows: 
\begin{align}\nonumber
&F_u(u)= \\\nonumber
&\small{\left(\text{col}(\ell^v_{u^v}(u_k^v,u_k^{-v}))_{k\in\bb{Z}_K, v\in\mc{V}}+\text{diag}(
2(\tilde{B}^{v})^\top \tilde{W}^v\tilde{B}^{v})_{v\in\mc{V}}\right)} u\\ \label{eq:FuDist}\
&\quad\quad+\text{col}((\tilde{B}^{v})^\top \tilde{w}^v )_{v\in\mc{V}}\\[5pt]
&F_x^v = 2(\tilde{B}^{v})^\top \tilde{W}^v \tilde{A}^v,\label{eq:FxDist}
\end{align}
%
%
%
where $\tilde{A}^v$ is the free response matrix of the local dynamics~\eqref{eq:DynDecomp} and $\tilde{B}^{v}$ is the impulse response matrix of agent $v$, defined similarly as in~\eqref{eq:ResponseMat}. Next, we define a local  storage function family $V^{1,v}_{\bar x}(x^v_t) = (\Delta x^v_t)^\top P^v~(\Delta x_t), \; \forall v\in \mc{V}$ and make use of the inequality~\eqref{eq:Diss1} characterizing the static nonlinearity $\phi(\cdot)$. From then on, the derivation of the distributed LMI conditions in~\eqref{eq:StabCondLocal} follows similar steps as in Theorem~\ref{thm:StabCond} and asymptotic stability for the decoupled dynamics~\eqref{eq:DynDecomp} and stage cost~\eqref{eq:CostDecomp} follow. 
%

\subsection{Proof of Corollary~\ref{cor:1DstabInequ}}
\label{sec:Proof1DIneq}
To prove Corollary~\ref{cor:1DstabInequ}, we analyze conditions under which 
\begin{align}\label{eq:Qclv}
Q_{cl}^v = \left[\begin{smallmatrix}
(A^v)^\top P^vA^v - P^v &  (A^v)^\top P^v\hat{B}^v\\
* &  (\hat{B}^v)^\top P^v \hat{B}^v\\
\end{smallmatrix} \right]
+ \eta^v\left[\begin{smallmatrix} 0 & -\frac{1}{2}(F_x^v)^\top \\
* & - \mu I \end{smallmatrix} \right]
\end{align}
in~\eqref{eq:StabCondLocal} is negative definite.
%
We begin by taking Schur complements of~\eqref{eq:Qclv}, the conditions
%
%
\begin{align*}
&-(\hat{B}^v)^\top P^v \hat{B}^v + \eta^v \mu I\succ0;\\[5pt]
& P^v -(A^v)^\top P^vA^v - (-(A^v)^\top P^v\hat{B}^v+ \frac{\eta^v}{2}(F_x^v)^\top)\\ &( (-\hat{B}^v)^\top P^v \hat{B}^v + \eta^v \mu I)^{-1}(-(A^v)^\top P^v\hat{B}^v+  \frac{\eta^v}{2}(F_x^v)^\top)^\top \succ 0
\end{align*}
%
%
%
are necessary and sufficient for $Q_{cl}^v \prec 0$ and for the choice of $P^v =  1 $ in the 1D case where $A^v \in \R$ and $B^v \in \R$:
%
{\small
\begin{subequations}\label{eq:Case1}
\begin{align} \label{eq:Case1a}
&\mu>\frac{P^v (B^v)^2}{\eta^v}> 0,\\
\label{eq:Case1b}
&\frac{\eta^v \mu (A^v)^2}{\eta^v\mu -   (B^v)^2} + \frac{\eta^v \mu (B^v)^2 (W^v)^2}{\mu}   \sum_{k=0}^{K-2}((k+1)(A^v)^k)^2 < 1.
\end{align}\end{subequations}} %
%
%
Thus, we have that if~\eqref{eq:Case1} holds the distributed LMI condition in~\eqref{eq:StabCondLocal} holds, ensuring stability of the local subsystem of agent $v$.

\end{document}